\documentclass[runningheads]{llncs}
\usepackage{graphicx}
\usepackage{amsmath}
\usepackage{booktabs}
\usepackage{array}
\usepackage{listings}
\usepackage{tikz}
\newcommand*\circled[1]{\tikz[baseline=(char.base)]{
            \node[shape=circle,draw,inner sep=1pt,font=\sffamily\footnotesize] (char) {\textbf{#1}};}}

\usepackage[normalem]{ulem}
\usepackage{wasysym}

\newcommand{\lmttfont}{\fontfamily{lmtt}\selectfont}

\usepackage{color}
\definecolor{comments}{rgb}{0.13,0.55,0.13}
\definecolor{background}{rgb}{0.94, 0.97, 1.0}
\definecolor{strings}{rgb}{0.63,0.125,0.94}
\usepackage{listings}

\usepackage[all=normal,floats,leading,charwidths,bibbreaks,mathspacing,wordspacing,tracking]{savetrees}

\usepackage{xcolor}
\usepackage{verbatim}

\begin{document}
\title{Towards Runtime Verification via Event Stream Processing in Cloud Computing Infrastructures}
%\thanks{Supported by organization x.}}
%
\titlerunning{Towards Runtime Verification via Event Stream Processing}
%\author{First Author\inst{1}\orcidID{0000-1111-2222-3333} %\and
%Second Author\inst{2,3}\orcidID{1111-2222-3333-4444} \and
%Third Author\inst{3}\orcidID{2222--3333-4444-5555}}
%\institute{Princeton University, Princeton NJ 08544, USA \and
%Springer Heidelberg, Tiergartenstr. 17, 69121 Heidelberg, Germany

\author{Domenico Cotroneo \and
Luigi De Simone \and
Pietro Liguori \and 
Roberto Natella \and 
\\Angela Scibelli
}

\authorrunning{D. Cotroneo et al.}
\institute{DIETI, University of Naples Federico II, Italy
%\and
\email{\{cotroneo,luigi.desimone,pietro.liguori,roberto.natella\}@unina.it
\\
%\and
%University of Naples Federico II, Italy\\
ang.scibelli@studenti.unina.it}
}

\maketitle              
% typeset the header of the contribution %
\begin{abstract}
Software bugs in cloud management systems often cause erratic behavior, hindering detection, and recovery of failures. As a consequence, the failures are not timely detected and notified, and can silently propagate through the system. 
To face these issues, we propose a lightweight approach to runtime verification, for monitoring and failure detection of cloud computing systems. 
We performed a preliminary evaluation of the proposed approach in the OpenStack cloud management platform, an ``off-the-shelf'' distributed system, showing that the approach can be applied with high failure detection coverage.
%The abstract should briefly summarize the contents of the paper in 15--250 words.

\end{abstract}
\section{Introduction}
\label{sec:intro}
Nowadays, the cloud infrastructures are considered a valuable opportunity for running services with high-reliability requirements, such as in the telecom and health-care domains \cite{dang2019survey,yin2015joint}. 
Unfortunately, residual software bugs in cloud management systems can potentially lead to high-severity failures, such as prolonged outages and data losses.
These failures are especially problematic when they are \emph{silent}, i.e., not accompanied by any explicit failure notification, such as API error codes, or error entries in the logs. This behavior hinders the timely detection and recovery, lets the failures to silently propagate through the system, and makes the traceback of the root cause more difficult, and recovery actions more costly (e.g., reverting a database state) \cite{cotroneo2019enhancing,cotroneo2019bad}.

To face these issues, more powerful means are needed to identify these failures at runtime. A key technique in this field is represented by \emph{runtime verification strategies}, which perform redundant, end-to-end checks (e.g., after service API calls) to assert whether the virtual resources are in a valid state. For example, these checks can be specified using temporal logic and synthesized in a runtime monitor \cite{delgado2004taxonomy,chen2007mop,zhou2014runtime,rabiser2017comparison}, e.g., a logical predicate for a traditional OS can assert that a thread suspended on a semaphore leads to the activation of another thread \cite{arlat2002dependability}. 
%In the context of cloud computing systems, the predicates should test at runtime the availability of virtual resources (e.g., volumes and connectivity).
Runtime verification is now a widely employed method, both in academia and industry, to achieve reliability and security properties in software systems \cite{bartocci2018lectures}. 
This method complements classical exhaustive verification techniques (e.g., model checking, theorem proving, etc.) and testing. %...by analyzing specific executions of a system and checks that the runtime behavior of the system complies with certain properties (e.g., freedom from data races and deadlocks), which are expressed by formal specifications. 

In this work, we propose a lightweight approach to runtime verification tailored for the monitoring and analysis of cloud computing systems. 
We used a non-intrusive form of tracing of events in the system under test, and we build a set of lightweight monitoring rules from correct executions of the system in order to specify the desired system behavior. 
We synthesize the rules in a runtime monitor that verifies whether the system’s behavior follows the desired one. Any runtime violation of the monitoring rules gives a timely notification to avoid undesired consequences, e.g., non-logged failures, non-fail-stop behavior, failure propagation across sub-systems, etc. 
Our approach does not require any knowledge about the internals of the system under test and it is especially suitable in the multi-tenant environments or when testers may not have a full and detailed understanding of the system.
We investigated the feasibility of our approach in the OpenStack cloud management platform, showing that the approach can be easily applied in the context of an ``off-the-shelf'' distributed system. 
In order to preliminary evaluate the approach, we executed a campaign of fault-injection experiments in OpenStack. Our experiments show that the approach can be applied in a cloud computing platform with high failure detection coverage.

In the following of this paper, Section~\ref{sec:related} discusses related work; Section~\ref{sec:approach} presents the approach; Section~\ref{sec:case_study} presents the case study; Section~\ref{sec:experiments} experimentally evaluates the approach;  Section~\ref{sec:conclusion} concludes the paper.

\section{Related Work}
\label{sec:related}
Promptly detecting failures at runtime is fundamental to stop failure propagation and mitigate its effects on the system. In this work, we exploit runtime verification to state the correctness of a system execution according to specific properties. In literature, some studies refer to runtime verification as runtime monitoring or dynamic analysis. Runtime monitoring consists of the observation of behaviors of the target system during its operation instead of verifying the system according to a specific model.

Over the last decades, several efforts have been spent on methodologies and tools for debugging and monitoring distributed systems. 
\textit{Aguilera et al.} \cite{aguilera2003performance} proposed an approach to collect black-box network traces of communications between nodes. The objective was to infer causal paths of the requests by tracing call pairs and by analyzing correlations.
Magpie \cite{barham2003magpie} and Pinpoint \cite{chen2004path} reconstruct causal paths by using a tracing mechanism to record events at the OS-level and the application server level. The tracing system tags the incoming requests with a unique \emph{path identifier} and links resource usage throughout the system with that identifier.
\textit{Gu at al.} \cite{gu2018kerep} proposes a methodology to extract knowledge on distributed system behavior of request processing without source code or prior knowledge. The authors construct the distributed system's component architecture in request processing and discover the heartbeat mechanisms of target distributed systems.
Pip \cite{reynolds2006pip} is a system for automatically checking the behavior of a distributed system against programmer-written expectations about the system. Pip provides a domain-specific expectations language for writing declarative descriptions of the expected behavior of large distributed systems and relies on user-written annotations of the source code of the system to gather events and to propagate path identifiers across chains of requests.
OSProfiler \cite{osprofiler} provides a lightweight but powerful library used by fundamental components in OpenStack cloud computing platform \cite{openstack}. OSProfiler provides annotation system that can be able to generate traces for requests flow (RPC and HTTP messages) between OpenStack subsystems. These traces can be extracted and used to build a tree of calls which can be valuable for debugging purposes. To use OSProfiler, it is required deep knowledge about OpenStack internals, making it hard to use in practice.

Research studies on runtime verification focused on formalisms for describing properties to be verified. Typically, a runtime verification system provides a Domain Specification Language (DSL) for the description of properties to be verified. The DSL can be a stand-alone language or embedded in an existing language. Specification languages for runtime verification can be regular, which includes temporal logic, regular expressions, and state machines, but also non-regular, which includes rule systems, stream languages. 

In the runtime verification literature, there is an established set of approaches for the specification of temporal properties, which include Linear Temporal Logic (LTL) \cite{pnueli1977temporal}, Property Specification Patterns (PSP) \cite{dwyer1999patterns}, and Event Processing Language (EPL) \cite{esper}.
Linear Temporal Logic is the most common family of specification languages. This approach supports logical and temporal operators. LTL is extensively used as specification language in many model checkers \cite{cimatti2002nusmv,blom2010ltsmin,holzmann1997model}.
The Property Specification Patterns consist of a set of recurring temporal patterns. Several approaches use PSP and/or extend original patterns used in \cite{bianculli2012specification}. %\cite{decserflow, aalst_pesic, elgammal, propols}.
Event Processing Language is used to translate event patterns in queries that trigger event listeners whether the pattern is observed in the event stream of a Complex Event Processing (CEP) environment \cite{wu2006high}. 
%In general, CEP is a technology for the collection, aggregation, and analysis of sequences of events that originated from various sources, occurring at different moments in time. 
The most interesting characteristic of CEP systems is that can be used in \textit{Stream-based Runtime Verification} or \textit{Stream Runtime Verification} (SRV) tools. SRV is a declarative formalism to express monitors using streams; the specifications are used to delineate the dependencies between streams of observations of the target systems and the output of the monitoring process. 
%SRV allows describing also a performance of quantitative/statistical analysis of traces.

%Lola \cite{d2005lola} is an SRV tool and implements a runtime verification as a stream computation, where output streams are defined in terms of input streams and/or other output streams. In particular, Lola defines a specification language and algorithms for both online and offline monitoring of synchronous systems and can be used to describe correctness/failure assertions but also statistical measures. 
%Esper \cite{esper} is an open-source software product for CEP and streaming analytics supporting Java and .NET languages. Esper provides an EPL language, a compiler, and a runtime environment. The language is declarative and data-oriented and extends the SQL-standard for analyzing streams of events with respect to time. 
%The Esper compiler compiles EPL source code into Java bytecode and the resulting executable code runs on a JVM within the Esper runtime environment. The Esper runtime provides an engine for online and real-time analysis. 
%Finally, Esper is designed to provide low latency and high throughput and to be lightweight in terms of memory, CPU, and IO usage.

In \cite{zhou2014runtime}, \textit{Zhou et al.} propose a runtime verification based trace-oriented monitoring framework for cloud computing systems. The requirements of the monitoring can be specified by formal specification language, i.e. LTL, Finite State Machine (FSM). The tracing adopted in this approach is fine-grained, in which traces are a collection of events and relationships: every event records the details of one execution step in handling the user request (function name, duration), every relationship records the causal relation between two events. Using both the events and the relationships, it is possible to represent a trace into a so-called \textit{trace tree}. In a trace tree, a node represents an event and an edge represents a relationship between events. This approach is generalizable at the cost of accessing the target source code to get the knowledge needed for instrumenting the code and gaining information about events relationships. However, this is not always the case, leading this approach difficult to exploit in practice. In \cite{power2019providing}, \textit{Power and Kotonya} propose Complex Patterns of Failure (CPoF), an approach that provides reactive and proactive Fault-Tolerance (FT) via Complex Event Processing and Machine Learning for IoT (Internet of Things). Reactive-FT support is used to train Machine Learning models that proactively handle imminent future occurrences of known errors. Even if CPoF is intended for IoT systems, it inspired us in the use of Complex Event Processing to build the monitor.

The proposed approach presents several points of novelty compared to state-of-the-art studies and tools in runtime verification literature. In particular, the proposed methodology relies on \textit{black-box tracing}, instead of regular tracing, avoiding knowing about system internals and the collection of information about the relationships between events (i.e., uncorrelated events). Further, we provide a new set of monitoring rules that well fit distributed systems and cloud computing infrastructure requirements, in which we need to face peculiar challenges like multi-tenancy, complex communication between subsystems, lack of knowledge of system internals. Based on the analysis of the events collected during system operation, we can specify the normal behavior of the target system and perform \textit{online anomaly detection}.

\section{Proposed Approach}
\label{sec:approach}
\begin{figure}[t]
    \centering
    \includegraphics[scale=0.3]{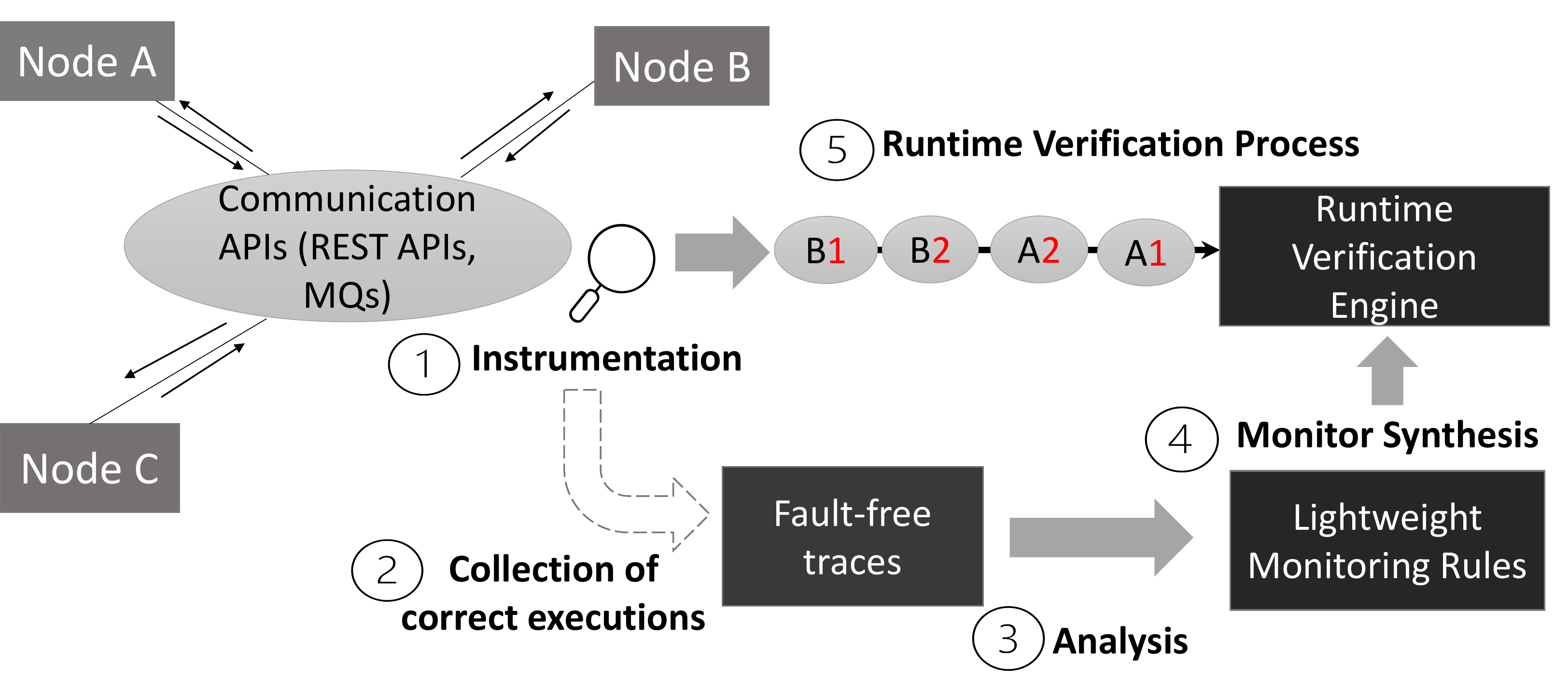}
    \vspace{-0.35cm}
    \caption{Overview of the proposed approach.}
    \label{fig:approach_overview}
    \vspace{-0.5cm}
\end{figure}

Figure \ref{fig:approach_overview} shows an overview of the proposed approach. 
Firstly, we instrument the system under test to collect the events exchanged in the system during the experiments (step \circled{1}). Our instrumentation is a form of \textit{black-box tracing} since we consider the distributed system as a set of black-box components interacting via public service interfaces. To instrument the system, we do not require any knowledge about the internals of the system under test, but only basic information about the communication APIs being used. This approach is especially suitable when testers may not have a full and detailed understanding of the entire cloud platform.
Differently from traditional distributed system tracing \cite{osprofiler}, this lightweight form of tracing does not leverage any propagation of the event \textit{IDs} to discriminate the events generated by different users or sessions.

In the step \circled{2}, we collect the \textit{correct executions} of the system. To define its normal (i.e., correct) behavior, we exercise the system in ``fault-free'' conditions, i.e., without injecting any faults.
Moreover, to take into account the variability of the system, we repeat several times the execution of the system, collecting different ``\textit{fault-free traces}'', one per each execution. We consider every fault-free trace a \textit{past correct execution} of the system.

Step \circled{3} analyzes the collected fault-free traces to define a set of \emph{failure monitoring rules}. These rules encode the expected, correct behavior of the system, and detect a failure if a violation occurs. This step consists of two main operations. Firstly, the approach extracts only the attributes useful for expressing the monitoring rules (e.g., the name of the method, the name of the target system, the timestamp of the event, etc.). 
Then, we define the failure monitoring rules by extracting ``\textit{patterns}'' in the event traces. We define a ``\textit{pattern}'' as a recurring sequence of (not necessarily consecutive) events, repeated in every fault-free trace, and associated with an operation triggered by a workload.
In this work, we identify patterns by manually inspecting the collected traces. In future work, we aim to develop algorithms to identify patterns using statistical analysis techniques, such as invariant analysis \cite{ernst2007daikon,yabandeh2011finding,grant2018inferring}.  

In general, we can express a monitoring rule by observing the events in the traces.
For example, suppose there is an event of a specific type, say \textit{A}, that is eventually followed by an event of a different type, say \textit{B}, in the same user session (i.e., same ID). The term \textit{event type} refers to all the events related to a specific API call. This rule can be translated into the following pseudo-formalism.

\begin{equation}
%\label{formula:single}
     a \to b\ and  \ id(a) = id(b), \ \ with \ a \in A, \  b \in B
\end{equation}

The rules can be applied in the multi-user scenario and concurrent systems as long as the information on the IDs is available. However, introducing an ID in distributed tracing requires both in-depth knowledge about the internals and intrusive instrumentation of the system. Therefore, to make our runtime verification approach easier to apply, we propose a set of coarse-grained monitoring rules (also known as \textit{lightweight monitoring rules}) that do not require the use of any propagation ID. 
To apply the rules in a multi-user scenario, we define two different sets of events, A and B, as in the following.
\begin{equation}
\begin{split}
A = \{all \ distinct \ events \ of \ type \ ``A" \ happened \ in \ [t,\ t + \Delta]\} \\
B = \{all \ distinct \ events \ of \ type \ ``B" \ happened \ in \ [t,\ t + \Delta]\}
\end{split}
\end{equation}

\noindent
with $\vert A \vert = \vert B \vert = n$. 
Our monitoring rule for the multi-user case then asserts that there should exist a binary relation $R$ over $A$ and $B$ such that:

\begin{equation}
\begin{split}
    R = \{ (a,b) \in A \times B ~|~ & a \to b, \\
    & \not\exists ~ a_i, a_j \in A, ~ b_k \in B ~|~ (a_i, b_k), (a_j, b_k), \\
    & \not\exists ~ b_i, b_j \in B, ~ a_k \in A ~|~ (a_k, b_i), (a_k, b_j)  ~ \}\\
\end{split}
\end{equation}

\noindent
with $i,j,k \in [1,n]$. That is, every event in $A$ has an event in $B$ that follows it, and every event $a$ is paired with exactly one event $b$, and viceversa. 
These rules are based on the observation that, if a group of users performs concurrent operations on shared cloud infrastructure, then a specific number of events of type A is eventually followed by the same number of events of type B. The idea is inspired by the concept of flow conservation in network flow problems. Without using a propagation ID, it is not possible to define the couple of events \(a_i\) and \(b_i\) referred to the same session or the same user \(i\), but it is possible to verify that the total number of events of type A is equal to the total number of events of type B in a pre-defined time window. 
We assume that the format of these rules can detect many of the failures that appear in cloud computing systems: if at least one of the rules is violated, then a failure occurred.

Finally, we synthesize a monitor from failure monitoring rules, expressed according to a specification language (step \circled{4}). 
The monitor takes as inputs the events related to the system under execution, and it checks, at runtime, whether the system's behavior follows the desired behavior specified in the monitoring rules (step \circled{5}). Any (runtime) violation of the defined rules alerts the system operator of the detection of a failure.

\section{Case Study}
\label{sec:case_study}
In this paper, we investigated the feasibility of the proposed approach in the context of a large-scale, industry-applied case study. 
In particular, we applied the approach in the OpenStack project, which is the basis for many commercial cloud management products \cite{OpenStackProducts} and is widespread among public cloud providers and private users \cite{OpenStackUsers}. Moreover, OpenStack is a representative real-world large software system, which includes several sub-systems and orchestrates them to deliver rich cloud computing services. 
The most fundamental services of OpenStack \cite{denton2015learning,solberg2017openstack} are (i) the \textbf{Nova} subsystem, which provides services for provisioning instances (VMs) and handling their life cycle; (ii) the \textbf{Cinder} subsystem, which provides services for managing block storage for instances; and (iii) the \textbf{Neutron} subsystem, which provides services for provisioning virtual networks for instances, including resources such as \emph{floating IPs}, \emph{ports} and \emph{subnets}. Each subsystem includes several components (e.g., the Nova sub-system includes \emph{nova-api}, \emph{nova-compute}, etc.), which interact through message queues internally to OpenStack. The Nova, Cinder, and Neutron sub-systems provide external REST API interfaces to cloud users.
To collect the messages (i.e., the events) exchanged in the system, we instrumented the \emph{OSLO Messaging library}, which uses a message queue library and it is used for communication among OpenStack subsystems, and the \emph{RESTful API libraries} of each OpenStack subsystem, which are used are used for communication between OpenStack and its clients.
In total, we instrumented only $5$ selected functions of these components (e.g., the {\lmttfont cast} method of OSLO to broadcast messages), by adding very simple annotations only at the beginning of these methods, for a total of 20 lines of code. We neither added any further instrumentation to the subsystems under test nor used any knowledge about OpenStack internals.

We collected one hundred correct executions by running the same workload in fault-free conditions. 
This workload configures a new virtual infrastructure from scratch, by stimulating all of the target subsystems (i.e., Nova, Neutron, and Cinder) in a balanced way. The workload creates VM instances, along with key pairs and a security group; attaches the instances to an existing volume; creates a virtual network consisting in a subnet and a virtual router; assigns a floating IP to connect the instances to the virtual network; reboots the instances, and then deletes them.
We implemented this workload by reusing integration test cases from the \emph{OpenStack Tempest} project \cite{openstack_tempest}, since these tests are already designed to trigger several subsystems and components of OpenStack and their virtual resources.

After the fault-free traces collection, we extract the information associated with every event within the trace. In particular, we record the time at which the communication API has been called and its duration, the component that invoked the API (\emph{message sender}), and the remote service that has been requested through the API call (\emph{called service}).
Internally, the approach associates an \textit{event name} to every collected event within a trace, so that two events of the same type are identified by the same name.
In particular, we assign a unique name to every distinct pair $<$\emph{message sender}, \emph{called service}$>$ (e.g., $<$\emph{Cinder}, \emph{attach volume}$>$).

\subsection{Monitoring Rules}

To determine the monitoring rules, we manually identified common patterns, in terms of events, in the fault-free traces. In particular, we determined a set of patterns for all the operations related to the workload execution (e.g., operations related to the instances, volumes, and networks).
 For example, the analysis of the events related to the attach of a volume to an instance pointed out common three different patterns in the fault-free traces. We derived failure monitoring rules for each pattern. Listing~\ref{lst:add_volume} shows three different monitoring rules, expressed in a pseudo-formalism, related to the {\lmttfont Volume Attachment} operation.

\begin{minipage}{\linewidth}
\begin{lstlisting}[caption={\lmttfont Volume Attachment} monitoring rules, label={lst:add_volume}]
Rule#1: event(name = "compute_reserve_block_device_name") is eventually followed by event( name = "compute_attach_volume")

Rule#2: event( name = "compute_attach_volume") is eventually followed by event( name = "cinder-volume.localhost.localdomain@lvm_initialize_connection")

Rule#3: Pattern of Rule#2 is eventually followed by event(name="cinder-volume.localhost.localdomain@lvm_attach_volume")
\end{lstlisting}
\end{minipage}

We derived the first rule by observing that, during the attachment of a volume, the {\lmttfont <compute, reserve\_block\_device\_name>} event is always followed by the {\lmttfont <compute, attach\_volume>} event in every fault-free trace. 
Indeed, to perform such an operation, the {\lmttfont reserve\_block\_device\_name} method, a synchronous RPC call, is called before the {\lmttfont attach\_volume} nova-compute API to get and reserve the next available device name. {\lmttfont Rule\#2} follows the same structure of the {\lmttfont Rule\#1}. {\lmttfont Rule\#3} shows the possibility to write more complex rules, involving more than just two events.

In the same way, we derived rules for all further operations related to the volumes and the instances. Instead, the identification of the rules for network operations is different and more complex. Indeed, network operations are performed by the Neutron sub-system in an asynchronous way, such as by exchanging periodic and concurrent status polls among agents deployed in the datacenter and the Neutron server. This behavior leads to more non-deterministic variations of the events in the traces. 
Given the high source of non-determinism affecting the network operations, it is not possible to create rules based on event ordering.  
Therefore, to find these patterns, we observed the repetitions of the Neutron-associated events both in the fault-free and the faulty traces (i.e., with a fault injected in the Neutron subsystem). We found that, when the injected fault experiences a failure in the Neutron component, some network-related events occurred a much higher number than their occurrences in the fault-free traces.

\begin{figure}[t]
    \centering
    \includegraphics[scale=0.3]{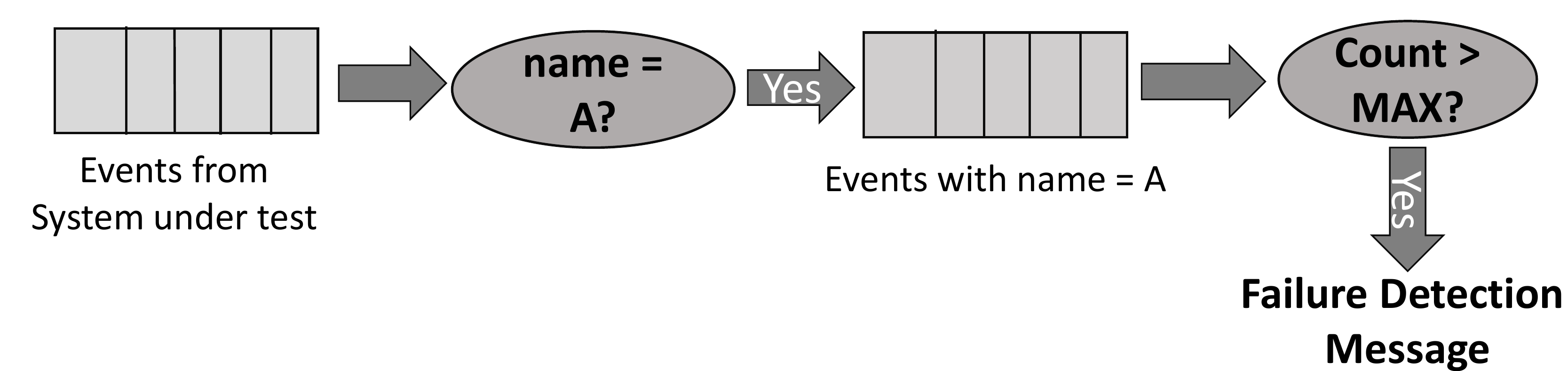}
    \vspace{-0.35cm}
    \caption{Example of Neutron SSH failure monitoring rules.}
    \label{fig:neutron_ssh_rules}
    \vspace{-0.5cm}
\end{figure}

For example, in several experiments targeting the Neutron component and that experienced a failure during the network operations, we found cases in which the event {\lmttfont <q-plugin, release\_dhcp\_port>} occurred more than $500$ times. However, analyzing all the fault-free executions, this specific event occurred at most $3$ times. 
Indeed, in such faulty experiments, the system repeatedly performed the same operation since it was unable to complete the request. 
Based on these observations, we defined the monitoring rules related to the network operations by checking, at runtime, if a specific event type occurred in a number higher than a threshold. Thus, for each event type, we defined a threshold as the higher number of times that the event type occurred during the fault-free executions. Fig. \ref{fig:neutron_ssh_rules} shows the logic adopted for the rules related to the network operations.

After identifying the rules, it is necessary to translate the rules in a particular specification language. We select EPL (\textit{Event Processing Language}) as specification language. EPL is a formal language provided by the Esper software \cite{esper}, that allows expressing different types of rules, such as temporal or statistical rules. It is a SQL-standard language with extensions, offering both typical SQL clauses (e.g., {\lmttfont select}, {\lmttfont from}, {\lmttfont where}) and additional clauses for event processing (e.g, {\lmttfont pattern}, {\lmttfont output}). In EPL, streams are the source of data and events are the basic unit of data. The typical SQL clause {\lmttfont insert into} is used to forward events to other streams for further downstream processing. We use the {\lmttfont insert into} clause for translating network operation rules using three interconnected statements, as shown in Listing \ref{lst:singleuser_neutron}.

\begin{minipage}{\linewidth}
\begin{lstlisting}[caption=EPL rules for the network operations,label={lst:singleuser_neutron}]
@name('S1') insert into EventNetworkStream select *
from Event where name='q-plugin_release_dhcp_port';
@name('S2') insert into countInfoStream select count(*) as count1 from EventNetworkStream;
@name('NetworkRule#1') select  * from countInfoStream where count1 > maxEvent1
output when OutputTriggerVar1 = true then set OutputTriggerVar1 = false;
\end{lstlisting}
\end{minipage}
The first statement ({\lmttfont S1}) extracts {\lmttfont <q-plugin, release\_dhcp\_port>} events and forwards them to the stream {\lmttfont NetworkEventStream}. The second statement ({\lmttfont S2}) counts the number of events in {\lmttfont NetworkEventStream} and passes this information to the stream {\lmttfont CountInfoStream}. Finally, the third statement  ({\lmttfont NetworkRule\#1}) produces an output if the value in {\lmttfont CountInfoStream} is bigger than the maximum value. To avoid that the third statement outputs anytime it receives a new {\lmttfont <q-plugin, release\_dhcp\_port>} event after the first output, we use the {\lmttfont OutputTriggerVar} boolean variable, initialized to true and set to false after the first time the rule is verified. 

The monitor synthesis is automatically performed once EPL rules are compiled. The Esper Runtime acts like a container for EPL statements which continuously executes the queries (expressed by the statements) against the data arriving as inputs.
For more detailed information on Esper, we refer the reader to the official documentation \cite{esperDoc}.

\subsection{Multi User Case}

We applied the EPL statements, derived from the monitoring rules, also in the multi-user scenario. Since we do not collect an event ID, we use a \textit{counter} to take into account multi-user operations. Indeed, we use the counter as an event ID to relate couples of events.
We associate a different counter to each event type: when an event of a specific type occurs, we increment its counter. In particular, the translation of rules described in Listing~\ref{lst:add_volume} uses the clause of {\lmttfont pattern}, useful for finding time relationships between events. Pattern expressions usually consist of filter expressions combined with pattern operators. We use the pattern operators {\lmttfont every}, {\lmttfont followed-by} ($\to$), and {\lmttfont timer:interval}. The operator {\lmttfont every} defines that every time a pattern subexpression connected to this operator turns true, the Esper Runtime starts a new active subexpression. Without this operator, the subexpression stops after the first time it becomes true. The operator $\to$ operates on events order, establishing that the right-hand expression is evaluated only after that the left-hand expression turns true. The operator {\lmttfont timer:interval} establishes the duration of the time-window during which to observe the arriving events (it starts after that the left-hand expression turns true).
The value of the counter is sent, along with the event name, to the Esper Runtime.  
Listing~\ref{lst:multiuser_cinder} shows the EPL translation of the rule {\lmttfont Rule\#1} in the multi-user case.

\begin{minipage}{\linewidth}
\begin{lstlisting}[caption=EPL rule in the multi-user scenario,label={lst:multiuser_cinder}]
@name('Rule#1') select * from pattern [every a = Event(name="compute_reserve_block_device_name") -> (timer:interval(secondsToWait seconds) and not b=Event(name="compute_attach_volume", countEvent = a.countEvent))];
\end{lstlisting}
\end{minipage}
Every time the Esper Runtime observes an event {\lmttfont <compute, reserve\_block\_device\_name>} with its counter value, it waits for the receive of the event {\lmttfont <compute, attach\_volume>} with the same counter value within a time window of {\lmttfont secondsToWait} seconds. If this condition is not verified, the approach generates a failure detection message.

\section{Preliminary Experiments}
\label{sec:experiments}
To preliminary evaluate our approach, we performed a campaign of fault injection experiments in the OpenStack platform.
In our experiments, we targeted OpenStack version 3.12.1 (release \emph{Pike}), deployed on Intel Xeon servers (E5-2630L v3 @ 1.80GHz) with 16 GB RAM, 150 GB of disk storage, and Linux CentOS v7.0, connected through a Gigabit Ethernet LAN.
In particular, our tool \cite{profipy} injected the following fault types:

\begin{itemize}

    \item \textbf{Throw exception}: An exception is raised on a method call, according to pre-defined, per-API list of exceptions;
    
    \item \textbf{Wrong return value}: A method returns an incorrect value. In particular, the returned value is corrupted according to its data type (e.g., we replace an object reference with a null reference, or replace an integer value with a negative one);
    
    \item \textbf{Wrong parameter value}: A method is called with an incorrect input parameter. Input parameters are corrupted according to the data type, as for the previous fault type;

    \item \textbf{Delay}: A method is blocked for a long time before returning a result to the caller. This fault can trigger timeout mechanisms inside OpenStack or can cause a stall.
    
\end{itemize}

Before every experiment, we clean-up any potential residual effect from the previous experiment, in order to ensure that the potential failure is only due to the current injected fault. To this end, we re-deploy the cloud management system, remove all temporary files and processes, and restore the OpenStack database to its initial state.

In-between calls to service APIs, our workload generator performs \emph{assertion checks} on the status of the virtual resources, in order to reveal failures of the cloud management system.  In particular, these checks assess the connectivity of the instances through SSH and query the OpenStack API to ensure that the status of the instances, volumes, and the network is consistent with the expectation of the tests. In the context of our methodology, assertion checks serve as \emph{ground truth} about the occurrence of failures during the experiments (i.e., a reference for evaluating the accuracy of the proposed approach). 

We evaluated our approach in terms of the \textit{failure detection coverage} (FDC), defined as the number of experiments identified as failed over the total number of experiments that experienced a failure.
We focused only on the experiments that experienced a failure, for a total of $481$ faulty traces, one per each fault-injection experiment. We define an experiment as failed if at least one API call returns an error (\textbf{API error}) or if there is at least one assertion check failure (\textbf{assertion check failure}). Also, to evaluate the most interesting cases, we focused on the experiments in which the target system was not able to timely notify the failure (i.e., failure notified with a long delay or not notified at all), as described in our previous work \cite{cotroneo2019bad}.

The coverage provided by our runtime verification approach is compared with the coverage provided by OpenStack API Errors by design. API Errors notifies the users that the system is not able to perform a request, thus they work as a failure detection mechanism. %An API Error makes the system transition in a safe state (fail-stop). 
Table~\ref{tab:comparison} shows the FDC of both approaches considering different failure cases (related to different operations). 
The results show that our approach is able to identify a failure in the $79.38\%$ of the failures, showing significantly better performance of the OpenStack failure coverage mechanism. 
In particular, the table highlights how our rules are able to identify failures that were never notified by the system (Instance Creation and SSH Connection). The RV approach shows lower performance only in the Volume Creation case failure: this suggests the need to add further monitoring rules or to improve the existing ones for this specific case.

\begin{table}
\vspace{-0.5cm}
\caption{Comparison with API Errors Coverage}
\label{tab:comparison}
\centering
\begin{tabular}{>{\centering\arraybackslash}p{3cm}>{\centering\arraybackslash}p{3.5cm}>{\centering\arraybackslash}p{3.5cm} }
\toprule
\textbf{Failure Case} & \textbf{OpenStack FDC \%} & \textbf{RV FDC \%} \\ 
\midrule
\textit{Volume Creation} & 29.67  & 28.57  \\
\textit{Volume Attachmen}t & 25.33 & 92.00  \\
\textit{Volume Deletion} & 100 & 100 \\
\textit{Instance Creation} & 0.00 & 90.96 \\
\textit{SSH Connection} & 0.00 & 38.46 \\
\midrule
\textit{\textbf{Total}} & \textbf{23.96} & \textbf{79.38}\\
\bottomrule
\end{tabular}
\vspace{-0.5cm}
\end{table}

We evaluated our approach also in a simulated multi-user scenario.
To simulate concurrent requests, $10$ traces ($5$ fault-free and $5$ faulty) are ``mixed-together'' by alternating the events of all the traces but without changing the relative order of the events within every single trace.
The faulty traces are related to the same failure type (e.g., Volume Creation).
For each failure type, we performed the analysis $30$ times by randomly choosing both the fault-free and the faulty traces.
Table~\ref{tab:multiusers} shows the average FDC and the standard deviation of our monitoring rules for all the failure volume cases. The preliminary results can be considered promising. However, the high standard deviation indicates that the average FDC is very sensitive to the randomity of the analyzed traces.

\begin{table}
\vspace{-0.5cm}
\caption{Average FDC in the multi-user scenario}
\label{tab:multiusers}
\centering
\begin{tabular}{>{\centering\arraybackslash}p{3cm}>{\centering\arraybackslash}p{4cm}>{\centering\arraybackslash}p{4cm} }
\toprule
\textbf{Failure Case}  & \textbf{Avg FDC \%}\\ 
\midrule 
\textit{Volume Creation}   & \(32.00 \mp 12.42\)   \\ 
\textit{Volume Attachment}   & \(45.33 \mp 13.82\) \\
\textit{Volume Deletion}  & \(36.00  \mp 12.20\) \\
\midrule
\textit{\textbf{Total}} & \(37.78 \mp 13.88\) \\
\bottomrule
\end{tabular}
\vspace{-1cm}
\end{table}

\section{Conclusion and Future Work}
\label{sec:conclusion}

In this paper, we propose an approach to runtime verification via stream processing in cloud computing infrastructures. We applied the proposed approach in the context of the OpenStack cloud computing platform, showing the feasibility of the approach in a large and complex ``off-the-shelf'' distributed system.
We performed a preliminary evaluation of the approach in the context of the fault-injection experiments. The approach shows promising results, both in the single-user and simulated multi-user cases.

Future work includes the development of algorithms able to automatically identify patterns using statistical analysis techniques, such as invariant analysis.
We also aim to conduct fault-injection campaigns by using a multi-tenant workload in order to perform an evaluation in a real multi-user scenario and to analyze the overhead introduced by the approach.

\section*{Acknowledgements}
This work has been supported by the COSMIC project, U-GOV 000010--PRD-2017-S-RUSSO\_001\_001.
\bibliographystyle{splncs04}
\bibliography{bibliography}

\end{document}